# L'utilité de l'"échelle opératique pour considérer des stratégies d'intelligence et de guerre économique

Stéphane Goria

Maître de conférences en sciences de l'information et de la communication

Université de Lorraine, Crem

Stephane.goria@univ-lorraine.fr

**Introduction**

Le XX⁠e siècle a vu apparaître un nouveau niveau de considération dans la manière d'envisager les opérations militaires. Ce dernier dit « opérationnel » ou « opératique » se situe entre les niveaux tactique et stratégique. Ce niveau d'échelle souvent ignoré est aussi qualifié « d'opérationnel » dans le lexique militaire, car ce niveau est aussi celui de la zone d'opération (c'est-à-dire d'engagements coordonnés regroupant sur un large espace des forces appartenant à des corps différents faisant face à d'autres forces « ennemies »). Cependant, comme ce dernier qualificatif est porteur d'ambiguïtés, puisqu'il peut être facilement confondu avec le niveau tactique, nous lui préférons comme d'autres auteurs (Coutau-Bégarie, 1999 ; Pénisson, 2013 ; Henninger, 2014 ; Barrier, 2019 ; Boulanger, 2020) celui d'« opératique ».

L'opératique, après quelques tâtonnements au XVIII⁠e et au XIX⁠e siècle, a véritablement été envisagée par les penseurs militaires du XX⁠e siècle dans le but de faciliter la transition entre les mises en œuvre tactiques et stratégiques. Il s'agissait en premier lieu d'apporter des solutions aux blocages et problèmes posés d'une part par l'apparition d'armes individuelles de plus en plus meurtrières et d'autre part l'accroissement important des zones de conflits. En fait, la transition qui semble aller de soi, dans un sens (du stratégique vers le tactique) comme dans l'autre (du tactique vers le stratégique) n'est pas si évidente à réaliser et mérite d'être pensée et associée à des moyens de mise en œuvre qui lui soient propres. Les militaires parlent dans le cadre de sa pratique d'« art opératif », car le changement de niveau d'échelle ne permet que de porter un regard selon une perspective différente. Toutefois, celle-ci permet l'identification de problèmes et de solutions associées qui sont difficiles à distinguer. Mais, une fois la perspective opératique envisagée, pour effectuer des actions relevant de ce niveau, il est nécessaire de disposer d'une organisation qui soit adaptée à leur mise en œuvre.

La problématique que nous abordons dans cet article est du même ordre. Il s'agit d'estimer ce que la perspective opératique peut apporter à la compréhension de problèmes et d'actions civiles relevant de l'intelligence ou de la guerre économique. De ce fait, nous considérons l'opératique comme un nouveau point de vue générant une pensée et des solutions particulières s'appuyant sur des systèmes considérés comme agiles, mais pouvant avoir atteint leurs limites. Nous supposons ainsi que ce qui est valable pour l'opératique militaire est transposable, au moins partiellement, aux domaines civils d'un point de vue informationnel ou économique. Cette réflexion nous semble originale, car nous n'avons pas trouvé qu'un seul écrit (Swinners et Briet, 1993) qui tentait de présenter de manière succincte ce que pouvait être l'opératique appliquée dans le domaine civil. En effet, nous avons questionné les moteurs de recherches *Google Scholar, Lens.org, Cairn* et *Web of Science*, sans d'obtenir d'autre résultat (nous avons employé sur la plateforme *Cairn* des expressions telles que : « opératique » + « entreprises », « art operatif » + « civil », « art opérationnel » + « marché », etc.).

Nous considérons l'échelon opératique comme une perspective permettant d'identifier et de donner du sens à des actions menées par une organisation au détriment d'une autre, via des domaines d'intervention différents, sur une durée de plusieurs mois à quelques années, afin d'atteindre un objectif stratégique. Nous proposons ici de questionner la manière dont l'opératique contribue au « coup d'œil » (O'Brien, 1991) nécessaire aux responsables concernés par une action d'ampleur à gérer. Pour ce faire, nous commencerons par apporter quelques éléments de réponses à la question de ce qu'est l'opératique militaire en termes de caractéristiques et de solutions envisagées pour son application. Ensuite, nous proposons de les transposer dans un cadre économique à partir d'un cas d'exemple : celui de l'acquisition de l'essentiel des activités de l'entreprise Alstom par l'entreprise *General Electric* (GE). Nous nous limiterons à cet exemple en prenant pour modèle d'interprétation une *timeline* annotée qui nous a semblé être un outil approprié pour raisonner à l'échelle opératique.

1. **Bref historique et fondamentaux de la pensée opératique militaire**

L'apparition d'un niveau de considération dans la manière d'envisager de façon théorique et pratique les opérations militaires entre les niveaux tactique et stratégique n'émergea pas de manière spontanée. Il fut le fruit de décennies de réflexions militaires répondant à des problèmes auxquels la pensée traditionnelle, séparée en niveau tactique et stratégique classique, avait du mal à apporter des réponses. *Le dictionnaire de la Pensée stratégique* (Géré, 1999), par exemple, nous informe que la tactique et son niveau de considération se distinguent de celui de grande tactique ou stratégique au cours du XVIII[e] siècle. La tactique n'est plus l'art de la guerre, mais « *l'art de combattre en présence de l'ennemi et de savoir faire combattre des hommes avec leurs matériels* » (Géré, 1999). La stratégie est alors positionnée vis-à-vis de la tactique comme l'art de gérer l'amont et l'aval de la tactique, c'est-à-dire la gestion de différents enjeux tactiques dans une vision d'ensemble. Toutefois dès le XIX[e] siècle, l'analyse de certaines campagnes napoléoniennes dont la manœuvre de Ulm, puis la d'autres de la guerre de Sécession (pensée russe) ou la guerre franco-prussienne de 1870-1871 (pensée allemande), posent des questions auxquelles la distinction tactique et stratégique ne semble pas permettre de bien expliquer ce qui s'est produit.

Par exemple, la manœuvre (dite « sur les arrières ») de Ulm a permis à Napoléon Bonaparte en octobre 1805, de surprendre une armée autrichienne commandée par Karl Mack et de l'enfermer dans la ville de Ulm. Si on résume cette manœuvre, elle contient quelques éléments essentiels de l'art opératif, comme un très grand espace d'actions pour l'époque, une durée d'environ 2 mois allant du 23 août (départ de l'armée impériale du camp de Boulogne) au 20 octobre (capitulation de l'armée de Mack enfermée dans l'enceinte de Ulm), l'usage de la désinformation et autres techniques de tromperie pour induire l'ennemi en erreur sur ces intentions et mouvements, la célérité et l'atteinte des moyens logistiques de l'adversaire bien au-delà de sa position. Elle peut être considérée comme repère en tant que date de naissance de l'art opératif (Durand, 2013). Les divisions constituées en corps d'armée de Bonaparte surprennent par leur vitesse de déplacement (d'abord par leur départ du camp de Boulogne), puis par le trajet qu'elles empruntent en passant en majorité par quelques routes parallèles plus au nord et à l'est de Ulm (entre Donauwörth, située à environ 80 km à l'est de Ulm, et Ingolstadt, située à environ 130 km à l'est de Ulm) tout en faisant croire que leur avancée se fera de manière plus directe (en empruntant des routes situées entre Baden Baden et Friegbourg ; figure 1). Par ce mouvement, l'armée napoléonienne coupe l'armée de Mack de ses routes d'approvisionnement et menace en même temps Munich tout en se couvrant plus au nord en prenant Nuremberg (Hodge, 2016, p. 14). Prise au dépourvu, après quelques affrontements, l'armée Autrichienne est contrainte de s'enfermer dans la ville où, encerclée, elle capitule le 17 octobre et se rend le 20.

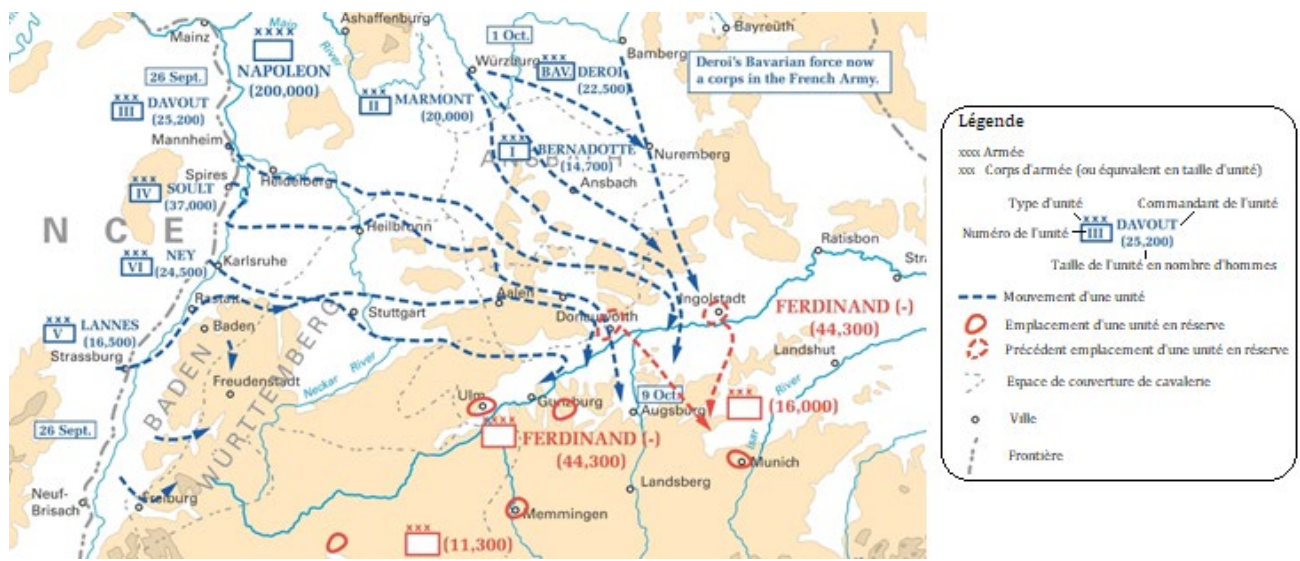

Figure 1. La manœuvre de Ulm (zoom d'après la carte dédiée de Westpoint.edu, 2022)

La défense française du 3 septembre 1870 au 26 janvier 1871, même si elle se conclut par une défaite française a aussi contribué à la réflexion militaire à propos de l'opératique. Si du point de vue des généraux et théoriciens français, il faudra attendre la fin de la Première Guerre mondiale pour voir émerger une forme de pensée opérative, du point de vue de l'état-major allemand, c'est autre chose. La résilience de la France et sa capacité à mobiliser et à envoyer au front de nouvelles armées rapidement pose question. Par exemple, si l'armée du maréchal François Achille Bazaine encerclée à Metz a capitulé assez tôt (27 octobre 1870), c'est plus pour des raisons politiques que militaires. Le chef d'état-major Helmuth von Motke va réfléchir longuement à cette campagne et à ce qui aurait pu se produire (notamment sans cette reddition rapide) pour revoir sa perception de la conduite des opérations militaires. Il va développer une certaine vision opérative sous forme de « direction opérationnelle » dont le rôle est de permettre la destruction de l'armée adverse en empêchant son soutien et ainsi briser la volonté de résistance de l'ennemi (Krause, 2005).

La pensée française de l'opératique se développe plus particulièrement durant la Première Guerre mondiale suite à l'impasse tactique sur le front de l'Ouest provoquée notamment par le perfectionnement des armes individuelles et de la mitrailleuse. La solution proposée par les unités blindées ne garantit pas encore de percée importante face à une défense organisée sur plusieurs lignes, surtout si le nombre de chars engagés et leur vitesse de progression sont assez faibles (c'est le cas pour l'offensive Nivelle dite du Chemin des Dames). L'approche opératique française va être développée sous la forme d'une succession de nombreuses « offensives limitées » proposées par le général Philippe Pétain (Doughty, 2005). Ainsi, fin 1918 après avoir repoussé une dernière offensive allemande, début août 1918, l'offensive des Cent-Jours commence. Les armées alliées attaquent les forces de l'axe sur un large front allant de la Picardie à la Meuse par un enchaînement d'attaques ponctuelles visant à percer les lignes adverses et à les déstabiliser au fur et à mesure. Les armées alliées bien coordonnées se comportent alors comme une équipe de rugby qui enchaîne les petites offensives et tentatives de percées en les menant sur toute la largeur du terrain par des décalages successifs. Une fois une offensive bloquée par l'adversaire qui a pu se regrouper, un déplacement latéral est effectué pour reprendre l'offensive un peu plus loin où l'adversaire dispose de moins de soutien (comme le ferait une équipe de rugby). Cette offensive a nécessité une organisation logistique et une coordination des assauts, mais a eu l'avantage de réduire les pertes humaines vis-à-vis des grandes offensives des années précédentes tout en provoquant un effondrement progressif de la défense allemande, l'amenant à sa capitulation.

Cependant, la véritable pensée opératique émerge en Russie Tsariste et continue d'évoluer durant la première décennie de l'Union soviétique où développe une pensée opératique moins centrée sur l'encerclement des armées adverses comme dans les réflexions allemandes. Elle se fonde, de manière distincte des autres pays, sur l'étude des raids de cavalerie de la guerre de Sécession qui ont parfois lieu sur des distances très importantes. Cette pensée évoluera aussi suite à la défaite contre le Japon en 1905 (Henninger, 2014), ainsi que par l'expérience des campagnes de la guerre civile russe menées sur des espaces très vastes (Kipp, 2005 ; Pénisson, 2013, p. 323). Dans l'approche soviétique, l'art opératif vise en premier lieu la désorganisation de l'ennemi considéré comme un système. Il nécessite une coordination d'armées sur de larges zones d'opération, des actions désinformation afin d'assurer un effet de surprise améliorant grandement les chances de réussite des offensives et des opérations en profondeurs pour s'assurer une progression importante visant l'atteinte d'objectifs stratégiques situés bien au-delà des premières lignes du front. Avec cette réflexion au niveau opératique, il est question de porter, à partir d'une première action d'une force importante, une percée même limitée au niveau tactique, puis, celle-ci obtenue, d'en profiter immédiatement pour débuter une nouvelle offensive prolongeant la première. Pour ce faire, il faut estimer le meilleur moment et terrain pour effectuer une percée majeure dans le dispositif ennemi par l'engagement d'une armée gardée en réserve. Celle-ci n'est pas chargée de gagner la bataille, mais de s'enfoncer sur les arrières de l'ennemi afin d'atteindre un objectif stratégique déterminé (isolement d'une armée, prise d'un carrefour logistique ou d'un autre élément vital pour le commandement ennemi, etc.). Cependant, les purges staliniennes des années 1930 vont reporter au milieu de l'année 1943 la mise en œuvre adéquate de l'opératique (Glantz, 2005).

Pour résumer, pour mener avec succès des actions relevant de l'art opératif moderne, il faut que les opérations envisagées bénéficient d'une préparation très importante, d'une bonne connaissance du terrain et des forces en présence, de l'élément de surprise autant que possible et que l'objectif stratégique visé soit clairement établi. Mais d'autres éléments sont encore plus essentiels à cette réussite : disposer de forces pouvant agir rapidement possédant une certaine autonomie pour s'adapter continuellement aux décisions adverses et autres facteurs imprévus, considérer un espace d'actions bien plus important que d'ordinaire, imprimer un rythme à ces actions pour garder l'initiative, définir un objectif stratégique situé bien au-delà de la ligne de front et assurer une préparation logistique conséquente. Ainsi, la célérité, le rythme imprimé et la coordination de nombreuses actions menées sur des zones distantes ont pour but de créer la confusion chez l'adversaire et alors faciliter l'atteinte de l'objectif stratégique. Cela implique que les forces employées soient agiles si l'on se réfère à la terminologie actuelle.

## 2. Transposition du domaine militaire aux civils

L'agilité est un adjectif à la mode qui ne s'applique pas qu'au domaine militaire. L'agilité peut être considérée comme la capacité dont une organisation dispose pour adapter rapidement ses processus afin d'agir ou de réagir rapidement en vue d'une menace ou d'une opportunité qu'elle a pu identifier (Barlerre, 2016). Dans le civil, l'agilité prend appui sur au moins deux interprétations propres au management des organisations. La première est focalisée sur l'agilité en tant que capacité relative à un évènement ou un adversaire à s'adapter pour conserver ou améliorer sa position. Il est possible d'associer à cette vision l'adaptation de la boucle OODA (Observation, Orientation, Décision, Action) proposée par John Boyd (Coram, 2002, p. 327-344). En effet, cette boucle est une modélisation d'un processus décisionnel dans le cadre d'un combat aérien qui a été adaptée à d'autres domaines (Ruggiero, 2015 ; Goria, 2019 ; Perkin & Abraham, 2021, p. 55). De plus, cette boucle dispose d'une transposition au niveau opératique (figure 2).

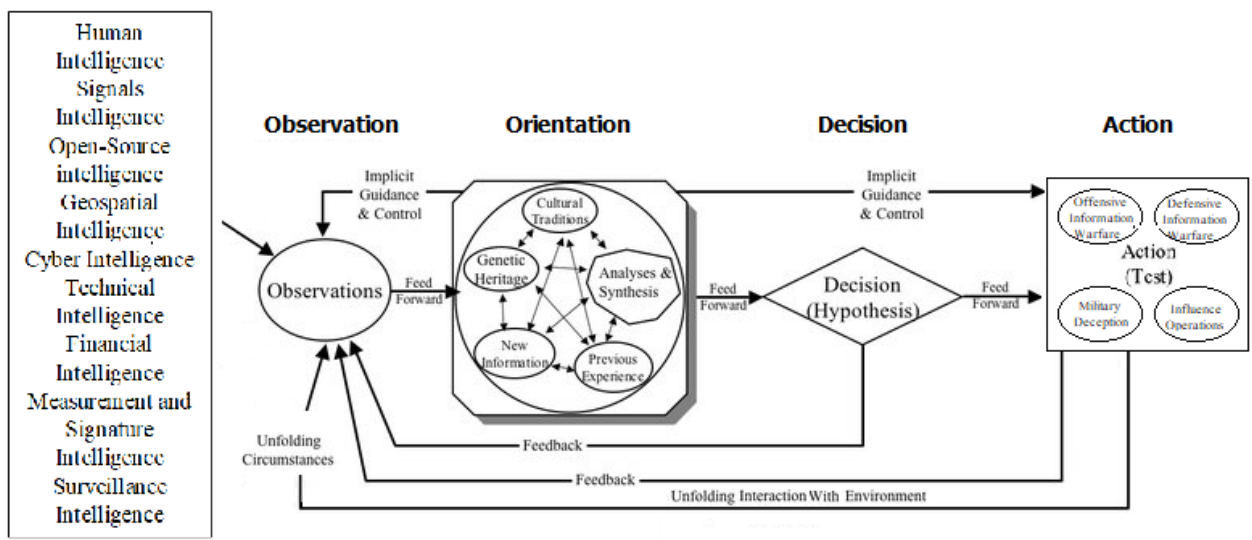

Figure 2. Modèle de boucle OODA pour l'opératique militaire (d'après Friedman, 2021, p. 79)

Avec ce modèle, l'agilité d'un système est exprimée relativement à un autre dans le contexte d'un affrontement où les actions de l'un influencent celles de l'autre. L'un a l'initiative et l'autre doit s'adapter au plus vite à la situation pour, autant que possible, reprendre l'initiative à son adversaire. Ainsi, celui qui a la boucle décisionnelle la plus efficace l'emporte sur l'autre. Cette vision de l'agilité comme capacité à répondre rapidement et de manière adéquate (et de préférence non prévisible) à un changement survenu dans l'environnement de manière à reprendre l'ascendant sur son adversaire a déjà fait l'objet d'interprétation dans le cadre d'applications au contexte des organisations, notamment dans des travaux récents portant sur la gestion informationnelle ou la sécurité informatique (Hoffman & Freyn, 2019 ; Moinet, 2019 ; Pech, 2019 ; Goria, 2020). Au niveau opératique, elle permet de garder en mémoire la relativité de l'agilité, sa dépendance à la célérité d'actions à mener de manière pertinente et non prévisible. Le modèle militaire de la boucle appliqué à l'échelle opératique et proposé par Chris Robinson (figure 2) reprend le modèle de Boyd en ajoutant un cadre aux actions d'information incluant les opérations d'influence, de déception (tromperie), de guerre de l'information offensive et défensive (Friedman, 2021, p. 79). Ce dernier modèle à l'avantage de permettre conceptuellement une transposition dans le domaine de l'entreprise assez simple, puisque tous les éléments proposés disposent d'un équivalent civil. Pour agir, l'organisation doit, par exemple, faire de la veille, analyser les informations collectées en fonction du moment et de sa culture, tenir compte des éléments adverses, comme de sa capacité en matière de guerre informationnelle (offensive et défensive), d'influence et de surprise.

La seconde approche de l'agilité est rattachée aux méthodes agiles et aux équipes qui les emploient. Les méthodes agiles trouvent leur origine dans les années 1980 et 1990 dans des travaux de gestion et d'une première initiative appelée *agile manufacturing* datant de 1991 (Barrand & Deglaine, 2013). Leur premier grand succès en termes de médiatisation date de 2001 lorsque de nombreux représentants de ces méthodes tentèrent de rassembler leurs approches et publièrent ce que l'on nomme désormais le manifeste agile (*agile manifesto*). Ce manifeste est toujours d'actualité et a eu un succès retentissant dans les développements de projets même au-delà du champ de l'informatique. Dans ce manifeste des valeurs et des principes d'agilité sont particulièrement mis en avant, dont :

- l'emploi préférable d'équipes réduites (de 5 à 20 membres) pour faciliter les interactions internes ;
- des développements séquencés en cycles de conception courts (de une à quatre semaines selon les types de projets) menant à des solutions fonctionnelles perfectibles qui peuvent ensuite être rapidement améliorées, complétées, assemblées ou même abandonnées ;
- des conceptions centrées d'abord sur une valeur à produire du point de vue du client et qui intègre avec bienveillance les modifications et changements du projet initial afin que lors de sa livraison la solution soit avant tout opérationnelle.

Si les succès de ces équipes sont nombreux et conviennent bien à une qualification d'agile, les choses se compliquent lorsqu'il s'agit de changer d'échelle pour passer d'équipes à des groupes plus importants, jusqu'à des entreprises agiles. Bien sûr, diverses méthodologies ont été proposées (*Scrum de Scrums*, *Scrum at Scale*, LeSS, Nexus, SAFe, Spotify, …). Certaines comme *Scrum de Scrums* (*Scrum of Scrums*) ou LeSS (*Large-Scale Scrum*) ne permettent que l'atteinte d'un niveau d'échelle supplémentaire. Ce passage à l'échelle supérieure encore assez simple peut être rapproché du modèle commandement militaire « équipe d'équipes » (*team of teams*) (McChrystal et al, 2015). L'idée est de transposer ce qui est faisable avec une équipe agile à un ensemble d'équipes plus ou moins importantes. Dans certains modèles, les échanges réguliers entre équipes se font via des ambassadeurs (comme dans *Scrum of*

*Scrums*), dans d'autres modèles (comme celui d'« équipe d'équipes ») chacun des membres d'une équipe se doit d'échanger régulièrement avec au moins un des membres des autres équipes. De plus, pour gagner en efficacité, les échanges doivent aussi continuer et être nombreux au sein de chaque équipe. Ainsi, même s'il existe une coordination centrale, les interactions entre équipes peuvent se faire plus rapidement, de même que la collecte et le partage d'informations. Les équipes correspondent dans ce cas aux divisions, corps d'armée ou armées modernes du domaine militaire puisqu'elles tentent à avoir des capacités similaires, c'est-à-dire être : plus rapide à réagir, capable de s'adapter rapidement à un changement décisionnel ou environnemental, relativement autonome tout en pouvant agir selon un schéma d'ensemble.

Dans cet esprit, l'échelon opératique permet d'interroger la dynamique de la circulation des informations et de la coordination des actions entre équipes ou cellules différentes. Nous y retrouvons la problématique posée par le général Stanley McChrystal (2015) à propos de la recherche d'une structure organisationnelle agile plus adaptée à un grand nombre d'individus. Par rapport aux modèles de commandement, il propose la solution d'une équipe d'équipe (sur la droite de la figure 3) qu'il trouve plus pertinente à condition que son application sur le terrain facilite véritablement le partage d'information, la coordination des actions et que chacun de ses acteurs dispose d'une vision réaliste de l'ensemble du circuit informationnel et décisionnel concerné. Dans ce modèle, il s'agit de créer plus d'interactions (même s'il est nécessaire d'imposer des espaces et des temps dédiés), de confiance mutuelle entre membres d'équipes différentes (en contraignant certaines personnes à se côtoyer) afin de faciliter en premier lieu la circulation des informations.

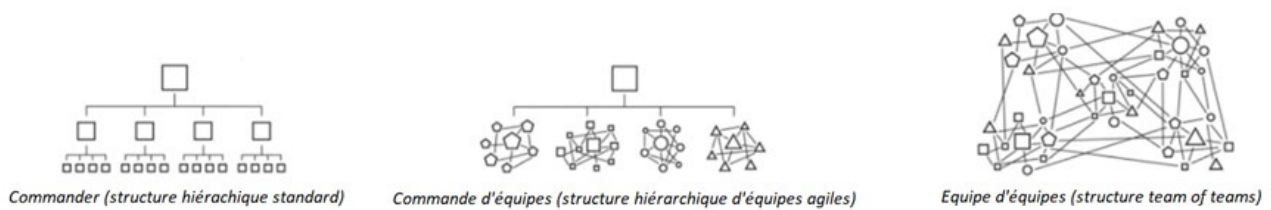

Figure 3. Trois structures d'organisation des personnels de la moins agile (à gauche) à la plus agile (à droite) (McChrystal et al, 2015, p. 204).

3. **L'opératique et la perception d'actions d'intelligence et de guerre économique**

L'opératique peut se traduire comme un ensemble de manœuvres plus ou moins importantes en termes de moyens alloués. Dans ce cadre, chaque manœuvre correspond à une action tactique dont la somme des effets doit permettre l'atteinte d'un objectif stratégique. Il nous semble que l'analogie d'un match rugby du point de vue d'une équipe permet de se faire une idée plus aisément de ce qui peut être fait, sachant que cette analogie est déjà employée dans certaines approches agiles (dont *Scrum* dont la traduction première est celle de mêlée). Certaines actions sont des leurres, une grande partie de la largeur du terrain est parcourue par le ballon (via de nombreuses passes) afin de fatiguer et tromper l'adversaire, de brèves tentatives de progressions par petits bons sont souvent tentées pour fixer l'attention et les forces adverses afin de se donner l'opportunité d'une tentative percée en profondeur visant un essai (objectif stratégique). De plus, au rugby, toutes les actions peuvent se transformer en percée si l'occasion se présente. Il faut que les joueurs concernés identifient le moment opportun, qu'ils puissent agir de manière coordonnée et non isolée (puisqu'un joueur seul n'arrive que très rarement à marquer un essai s'il ne bénéficie pas du soutien d'autres membres de son équipe). De même, ils ne doivent pas être seulement tournés vers l'offensive, mais aussi être capables de décrypter, de s'adapter aux actions adverses lorsqu'ils ont perdu l'initiative et tenter de la reprendre.

Comme nous l'avons précédemment évoqué, il y a eu très peu de tentatives pour proposer une adaptation de l'opératique en dehors du champ militaire. Dans une approche de marketing inspirée des approches militaires, nommée *Warketing*, Jean-Louis Swinners et Jean-Michel Briet (Swinners et Briet, 1993) ont défini l'opératique comme l'art de mener un plan en tant que contre plan, c'est-à-dire comme un ensemble d'actions perturbant et surprenant la stratégie adverse. Si nous ne nions pas le fait qu'il faut anticiper au moins une partie des actions et réactions adverses lors de l'élaboration d'un plan, cette définition nous a semblé encore trop vague pour pouvoir être bien comprise. Si ces auteurs n'approfondissent pas plus l'opératique, ils proposent tout de même l'emploi d'un planning annuel (*ano-planing*) des opérations afin de percevoir si un schéma d'ensemble est identifiable. Nous proposons de le remplacer par une *timeline* annotée permettant de considérer en plus les actions menées par catégorie afin de faciliter la vision d'ensemble.

La *timeline* elle-même est assez facile à réaliser. Bien entendu, ce sont les informations qui y figurent qui sont plus complexes à collecter. Afin de montrer les possibilités que cet outil offre, nous avons choisi de prendre pour exemple le rachat de la majorité des activités d'Alstom par le conglomérat Général Electric (GE). Cet exemple nous semble intéressant, car il relève d'actions de guerre économique entre deux grandes entreprises chacune appuyée de façon différentes par un état. De plus, il a déjà fait l'objet d'au moins une étude sous la perspective de la boucle OODA (Coussi & Moinet, 2019), mais pas sous l'angle opératique. Ainsi, nous proposons une *timeline* annotée à l'aide afin de rendre compte notamment d'une variété d'attaques relativement coordonnées :

- Des évènements et faits en faveur plutôt de GE ou de l'état américain (indiqués par le drapeau américain) d'une part, et de Alstom ou de l'état français (indiqués par le drapeau français) d'autre part ;
- Des actions d'influence illégales soupçonnées qui auraient contribué à la réussite du projet de GE (indiqués par des triangles chapeautés d'une flèche) ;
- Des évènements et faits complémentaires à la compréhension de l'histoire.

La durée des actions menées par GE pour réussir l'acquisition de l'essentiel des activités d'Alstom (son concurrent le plus important pour ses activités de production de turbines de gaz) est difficile à bien cerner. Nous la faisons débuter en janvier 2014 avec le dessin d'un segment central noir, mais elle peut être avancée d'un an, ce que nous représentons par des pointillés en amont. De même, si l'achat d'Alstom par GE est effectif le 2 novembre 2015 (fin du segment), l'histoire ne s'arrête pas complètement à ce moment, c'est pourquoi nous avons prolongé aussi ce segment par des pointillés.

Toutes les informations ne sont pas présentées sur cette *timeline*. Il s'agit seulement de mettre en perspective une partie de ces dernières et de voir comment un ensemble d'actions enchaînées sous différents axes ont permis à GE d'atteindre son objectif. Cette *timeline* est centrée sur des informations jugées importantes à l'échelon opératique. À l'échelon stratégique, le début de l'affaire se situerait bien plutôt, au moins vers 2004 avec les premiers accords de coopération entre Alstom et la Chine pour la production de turbines à vapeur (Alstom, 2004). C'est à ce moment que peut se situer le *casus belli* avec Alstom d'une part, et GE et l'état américain d'autre part. Le plan de guerre de GE s'appuie ensuite sur un dispositif américain de poursuites judiciaires extraterritoriales qui nécessite d'abord d'identifier un délit potentiel de corruption. Ainsi, la question qui nous intéresse est comment (une fois une faille en matière de corruption identifiée dans le dispositif d'Alstom) s'enchaînent et s'organisent les attaques combinées du *Département Of Justice* (DOJ) américain et de GE contre Alstom. Ces actions débutent seulement en 2010 à la suite d'une affaire de corruption datant de 2003 (Coussi & Moinet, 2019), mais à l'échelle opératique le démarrage de l'offensive se situe plutôt en 2011 avec l'inculpation de l'entreprise Alstom pour des faits de corruption.

La première victoire (percée du front Alstom) de GE a lieu le 23 avril 2014 lors d'une réunion aux USA où Patrick Kron PDG d'Alstom accepte de vendre son entreprise à GE. En amont de cette décision, il y a les poursuites, débutées en 2011, du DOJ américain auprès d'Alstom pour des faits de corruption en Indonésie. Ces faits relèvent de l'interprétation extraterritoriale du droit américain, mais sont suivis de plusieurs évènements dont la condamnation d'autres entreprises dont Total en juin 2013 pour 398 millions de dollars et une série d'arrestations de cadres importants d'Alstom en 2013 (dont une le jour même de l'entretien du 23 avril 2014 entre Patrick Kron et Jeffrey Immelt PDG de GE). À ce moment, il nous semble possible de considérer que Patrick Kron (contraint notamment par la perspective d'une amende très importante pour Alstom et de son incarcération aux USA pour plusieurs années) passe dans le camp de GE. En termes d'influence, c'est une étape importante, car désormais, seul l'état français, à partir de députés et de membres du gouvernement de l'époque, est en mesure de contrarier cette acquisition. Or, une partie du réseau d'influence d'Alstom et de Patrick Kron, à partir de ce moment, peut être considérée comme alignée sur le réseau d'influence de GE. Cet aspect peut expliquer en partie l'échec rencontré par les premières demandes de création d'une commission d'enquête à propos de cette tentative d'acquisition. Avec un regard analogique tout relatif, nous pouvons faire un parallèle entre cette situation économique et celle militaire de la défaite de Sedan (2 septembre 1870) qui a changé la situation et la direction militaires de la France face aux armées prussiennes sans mettre pour autant fin à la guerre.

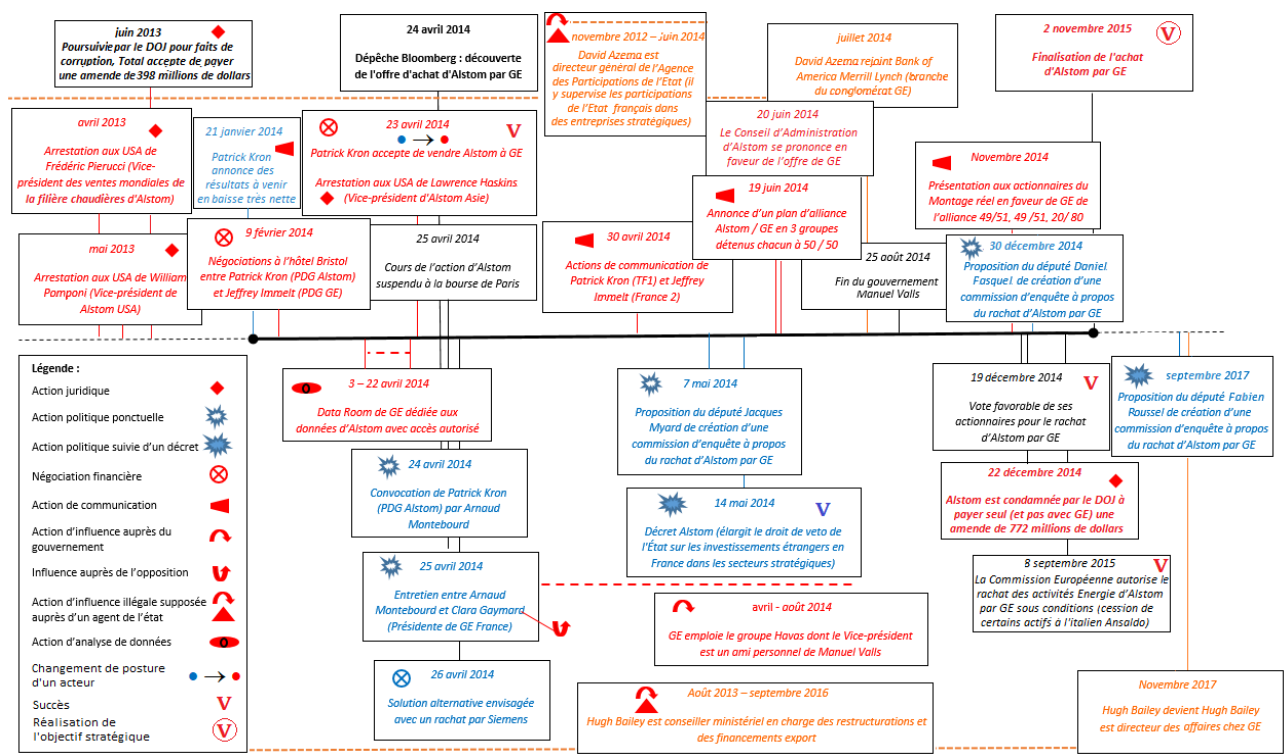

Figure 4. *Timeline* du rachat d'Alstom par GE

Au-delà de ces éléments, la *timeline* permet de voir qu'à partir de la mi-mai jusque fin décembre, l'initiative est clairement du côté de GE qui multiplie les négociations et montages financiers pour valoriser son projet. Toutefois, a priori certains obstacles peuvent se dresser devant la tentative d'acquisition (percée en profondeur) de GE : le décret n° 2014-479 « *relatif aux investissements étrangers soumis à autorisation préalable* » (donnant un droit de veto de l'état sur l'achat d'entreprises françaises jugées « stratégiques » pour garantir l'indépendance du pays), les demandes de création d'une commission d'enquête à propos de ce projet d'achat, un blocage potentiel des syndicats d'Alstom ou de ses actionnaires. Le décret, dit désormais « Alstom », pousse à l'élaboration du montage financier impliquant une participation de l'état français, mais n'empêche pas GE de devenir actionnaire majoritaire de chacune des trois entités d'Alstom (Énergie, Nucléaire, et les réseaux hydrauliques). Les actionnaires sont satisfaits de la proposition de versement de dividendes de la vente et les syndicats semblent en partie rassurés par les annonces de GE de maintien et de création de nouveaux emplois (AFP, 2014). Dans un premier temps, le réseau d'influence GE + Alstom semble avoir retardé la création d'une commission d'enquête, tandis que l'avis positif du gouvernement sur ce rachat a pu avoir été en partie influencé par certains de ses conseillers, dont David Azema (Quatrepoint, 2015, p. 165) et Hugh Bailey (de Colnet, 2021, p. 181) qui ont été ensuite employés à des postes importants chez GE ou l'une de ses filiales. Dans une démarche stratégique visant l'anticipation d'actions (conséquences) possibles, nous aurions pu ajouter à la figure 4, la pression exercée par GE, une fois le rachat d'Alstom effectué, sur l'entreprise EDF. En effet, une fois l'acquisition réalisée, GE « *devient de fait l'unique fournisseur pour l'entretien et la fourniture intégrale des pièces de rechange des cinquante-huit turbines Arabelle* » du parc nucléaire français (de Colnet, p. 19). Une grève de maintenance est ainsi organisée par GE en février 2016 afin de réduire ses coûts et risques auprès d'EDF qui finira par céder. Un grand nombre d'incidents en résulteront qui ont des résonnances encore actuelles concernant la production d'énergie électrique de la France.

Dans un cadre de veille relatif à une situation similaire, ce type de *timeline* peut être élaborée en pleine crise. Pour ce faire, nous proposons de remonter le fil des actualités propres à la situation en cours sur une année. Il s'agit d'y faire figurer des actions de natures différentes liées à l'entreprise cible, mais aussi de l'adversaire identifié. Ceci fait, les éléments d'influence peuvent y être ajoutés et le fil du temps remonté encore de quelques mois, si nécessaire. À ce stade, la *timeline* se transforme en tableau de bord de suivi des différents éléments en cours et permet d'élaborer de nouvelles hypothèses et stratégies sur ce qui pourrait arriver. Bien entendu, un dispositif de veille stratégique adéquat devrait débuter cette *timeline* bien plus tôt en identifiant en amont des signaux d'alerte indiquant une menace. Par exemple dans le cas d'Alstom, les premières arrestations de ces cadres dirigeants aux USA (avril et mai) auraient dû enclencher un état d'alerte, sachant que GE était bien identifiée comme l'un de ses principaux concurrents et que cette entreprise s'était déjà appuyée sur des poursuites judiciaires extraterritoriales liées au *Foreign Corrupt Practices Act* (FCPA) pour acheter des entreprises européennes comme Amersham (en 2004), Nycomed (en 2004) et Vetco Gray (en 2007) (Quatrepoint, 2015, p. 72). De plus, ces premières actions datent de la même époque que les accords de coopération entre Alstom et la Chine pour la production de turbines à vapeur. Les neuves années séparant ces accords des premières arrestations de cadres d'Alstom ont dû endormir la vigilance des responsables de cette entreprise ou les conforter dans leur fausse impunité à l'égard des lois américaines. Du point de vue de l'État français, dans ce cadre il y a nécessité, d'abord, une veille à l'échelon stratégique visant à être attentif aux procédures pénales relevant de l'extraterritorialité américaine (dont le FCPA ou les lois d'Amato-Kennedy et Helms-Burton concernant les pays sous embargo américain) portant sur des entreprises nationales. Puis, de la mise en place d'une autre veille au niveau opératique rapportant si des actions relevant de cet échelon sont menées à l'intention d'une entreprise spécifique. Cependant, une veille de cet ordre ne peut être utile que si la défense correspondante est organisée dans une vision d'ensemble (si le risque d'agression relève de l'opératique) et ne se limite pas, de fait, à l'action d'un ministère ou service unique.

**Conclusion**

L'exploitation de l'échelle opératique n'est ni intuitive ni aisée à mettre en œuvre, mais le fait de s'intéresser aux démarches informationnelles et aux stratégies des organisations selon cette dernière permet a priori de poser de nouvelles questions et de donner du sens à une série d'évènements. Comme dans notre cas d'exemple, l'apport de ce regard plus large via une *timeline* permet la prise d'un certain recul sur des évènements en cours ou passés, mais aussi le questionnement de correspondances entre des faits portant sur des sites géographiquement éloignés, des domaines d'activités différents, etc. Sous cette perspective, un veilleur ou un analyste envisageant des faits sous une perspective opératique se doit d'établir des corrélations, de faire des choix concernant la mise en avant de certaines informations par rapport à d'autres. Le veilleur est à ce moment dans le rôle d'un enquêteur travaillant sur différents dossiers qui peuvent potentiellement être reliés selon une succession de moments. Dès lors, si un schéma d'ensemble se dessine il pourra y avoir déclenchement d'une alerte, surtout si d'un point de vue stratégique certains signaux auront déjà pu être reconnus.

Cette proposition de transposition d'une conception militaire vers le civil possède quelques limites. La transposition étant du registre de l'analogie, elle nécessite une adaptation et ce genre d'opération n'est pas toujours réalisable sans une certaine déformation qui peut rendre le résultat inapproprié à l'usage visé. Le cas d'exemple choisi pour illustrer son apport potentiel, puisqu'il est unique, permet seulement d'entrevoir ce que le niveau opératique permet de faire apparaître, sans mettre en évidence une réelle méthodologie. De plus, les entreprises concernées doivent être de grande taille, ce qui réduit le champ d'application.

Tournée vers l'environnement de l'organisation, l'échelle opératique permet d'envisager l'atteinte d'un objectif stratégique à partir de la coordination de diverses actions tactiques (veille, influence, sécurité des informations, marketing, démarches juridiques, etc.). Ainsi, elle semble bien adaptée pour piloter des démarches d'IE en y ajoutant une dimension temporelle où les actions réalisées le sont selon un rythme et un schéma d'un ensemble. Tournée vers l'interne, l'opératique permet de considérer les verrous, goulot d'étranglement et freins dans le partage d'informations entre différentes cellules et strates d'une organisation. Nous pouvons attendre de cette perspective qu'elle permette de voir, à la fois, s'il y a un réel lien effectif entre les actions et les informations de terrain avec les informations et décisions stratégiques.

Des travaux de recherches portant sur la perspective opératique civile sont encore à mener. La littérature consacrée est presque qu'inexistante alors que les travaux scientifiques qui s'y intéressent d'un point de vue militaire et/ou historique sont un peu plus conséquents (une requête avec « art opératif », moins ambigu seul que « opératique » qui a un homonyme relatif à l'opéra, donne sur la plateforme *Cairn* 63 réponses, tandis que « operational art » donne 196 réponses sur le *Web of Science*). Certaines recherches futures devraient porter sur les méthodes et outils permettant de suivre ces flux décisionnels ou informationnels selon cette perspective. Ces outils peuvent être considérés, à partir du moment où les échelons tactique et stratégique existent au sein d'une organisation dans la mise en œuvre de l'une de ses démarches au-delà de simples éléments de langage. En ce sens, des actions de veille, de communication ou d'innovation peuvent être interrogées d'un point de vue de l'opératique pour savoir si cet échelon intermédiaire permet de clarifier certains faits ou problèmes. L'utilité de l'opératique mérite aussi d'être questionnée sur ce qu'elle apporte comme complément d'information pour comparer deux organisations, notamment dans l'analyse dynamique d'actions d'IE. Enfin, un questionnement devrait porter sur la réception même de cette conception d'une échelle intermédiaire entre les niveaux tactique et stratégique, de même que la compréhension ce que peut être une agilité opératique vis-à-vis d'une agilité stratégique.

## Bibliographie